\documentclass[aps, preprint, onecolumn,netbib]{revtex4} 
\usepackage{graphics}      % standard graphics specifications
\usepackage{graphicx}      % alternative graphics specifications
\usepackage{longtable}     % helps with long table options
\usepackage{url}           % for on-line citations
\usepackage{bm,color}            % special 'bold-math' package
\usepackage{amssymb, amsmath, enumerate, theorem, epsfig,subfigure,setspace}

\newcommand{\nb}[1]{\textcolor{black}{#1}}
\newcommand{\kk}[1]{\textcolor{black}{#1}}

\begin{document}
%\setstretch{2}

\title{Exotic states of matter with polariton chains}

\author{Kirill Kalinin$^1$, Pavlos G. Lagoudakis$^{1,2}$ and Natalia G. Berloff$^{1,3}$ }
\email[correspondence address: ]{N.G.Berloff@damtp.cam.ac.uk}
\affiliation{$^1$Skolkovo Institute of Science and Technology Novaya St., 100, Skolkovo 143025, Russian Federation}
\affiliation{$^2$Department of Physics and  Astronomy, University of Southampton, Southampton, SO17 1BJ, United Kingdom}
\affiliation{$^3$Department of Applied Mathematics and Theoretical Physics, University of Cambridge, Cambridge CB3 0WA, United Kingdom }

\date{\today}

\begin{abstract}{We consider linear periodic chains  of  exciton-polariton condensates formed by \kk{pumping polaritons non-resonantly}  into a \nb{linear network}. To the leading order such a sequence of condensates establishes relative phases  as to minimize  a classical  one-dimensional XY Hamiltonian with nearest and next to nearest neighbours. 
%From the complex Ginsburg-Landau model of the condensate we find analytical estimates of the coupling strengths between the condensates. 
We show that the low energy states of polaritonic linear chains   demonstrate  various classical regimes: ferromagnetic, antiferromagnetic and frustrated spiral phases, \nb{where  quantum or thermal fluctuations  are expected to give rise to a spin liquid state.} At the same time  nonlinear interactions at  higher pumping intensities \nb{bring about phase chaos} and  novel exotic phases.  }

\end{abstract}

\maketitle

Frustrated spin systems represent one of the most demanding problems of condensed matter physics \cite{Sachdev_NatPhys2008,spins}. Various strongly correlated states   are realised by 
Hubbard models \cite{hubbard,Auerbach1994}, that, in particular limits, reduce to
 spin models. Geometric frustration leads to a rich variety of possible spin configurations in the ground and excited states of these systems  due to 
 the competition between interactions and the geometry of the spin lattice with a potential of creating novel, exotic classical and quantum
phases, such as resonating valence bond states and valence bond solids, as well as 
various kinds of quantum, topological and critical  spin liquids  \cite{spinliquids,Alet2006}.  A spin liquid is a much sought  state of matter with applications  to high-temperature superconductivity and  quantum computation. In this state,  spins fluctuate in a liquid form without ever solidifying even in the ground state  \cite{balents}. In real systems, the classical spin Heisenberg Hamiltonians  are at  best  approximate. The real Hamiltonians are affected by \kk{assorted} perturbations  such as  quantum and thermal fluctuations, anisotropies,  disorder,  dipolar interactions, coupling to lattice degrees of freedom.  Depending on interactions  and connectivity classical Heisenberg models may exhibit frustration with a large ground state degeneracy. The ground state manifold has no energy scale of its own and therefore such perturbations  may bring about unusual  disordered spin liquid, Bose metal states with exotic  excitations and novel  phase transitions.  The
theoretical possibility of quantum fluctuations combined with geometric spin frustration  has been hotly debated since Anderson
proposed them in 1973 \cite{anderson} and only  recently have been realised experimentally \cite{han}. It remains to be seen which exotic phases  other types of perturbations can \kk{produce}. This can be elucidated by studying emergence phenomena: the dynamical generation of new types of degrees of freedom using 
 some  condensed matter systems to mimic (simulate) spontaneously and collectively different ones, possibly unknown or otherwise unrealizable.

Various systems have been proposed to act as analog classical or quantum simulators   to mimic condensed
matter phenomena and   realise \kk{disparate}  kinds of spin models: ultracold atomic and molecular
gases in optical lattices \cite{reviewUltracold,saffman,struck11,simon11},  defects and vacancies in semiconductors or dielectric materials \cite{pla,hanson}, magnetic impurities embedded in solid helium \cite{lemeshko13},   photons \cite{northup}, trapped ions \cite{kim10,lanyon}, superconducting q-bits \cite{corcoles},  network of optical parametric oscillators (OPOs) \cite{yamamoto11, yamamoto14}, coupled lasers \cite{coupledlaser}. 

Recently, we introduced polariton graphs as a new platform to study \kk{unconventional superfluids, spin liquids, and many other systems based on the $XY$ Hamiltonian} \cite{natmat17}.  Polariton condensates can be imprinted into any two-dimensional graph  by spatial modulation of the pumping laser \cite{tosi12} and can be easily scalable  to a large number of lattice sites. In the regime of non-resonant excitation\kk{,} the individual condensates in a polariton graph select their relative phases without the influence of the pumping laser. The condensation  is driven by bosonic stimulation, so that polariton graph condenses at the state with the phase-configuration that carries the highest polariton occupancy \cite{ohadi16}, \nb{which corresponds to the global minimum of the XY Hamiltonian. The structure of the XY Hamiltonian is set by the interaction strengths among the condensates that are defined by the systems parameters such as the graph geometry, the pumping profile and intensity.} The XY model is a mathematical abstraction of spin system such as 
a  disordered magnetic material composed
of competitively interacting spins. For such a system, a spin indexed by $i$ is represented by a two-dimensional unit vector ${\bf s}_i=(\cos\theta_i, \sin\theta_i)$, the energy is expressed by the XY Hamiltonian 
$
H_{XY}=-\sum_{ i<j} J_{ij}{\bf s}_i\cdot{\bf s}_j = - \sum_{ i<j} J_{ij}\cos(\theta_i-\theta_j),
$
 where $J_{ij}$ denotes the coupling coefficient between the $i$th and
$j$th spins. We refer to the coupling as ferromagnetic (antiferromagnetic) if $J_{ij}>0$ ($J_{ij}<0$). The XY model is a limiting case of the Heisenberg spin model  or $n$-vector model for $n=2$. The vectors correspond to the directions of spins (originally quantum mechanical) in a material in which the $z$-component of spins couples less than the $x$ and $y$ components. 
 Since $H_{XY}$  is the simplest model that undergoes the $U(1)$ symmetry-breaking transition, it has been discussed in the connection to classical magnetic systems, unconventional superfluid,  topological quantum information processing and storage,  quantum phase transitions,  spin-liquid and spin-ice phases and high-$T_c$  superconductivity.  

In this letter, we propose to exploit the properties of polariton graphs to generate various frustrated states and spinor phases. In particular, we show that at the condensation threshold depending on the structure of the corresponding XY Hamiltonian the  polariton condensates establish \nb{the relative phases that correspond to the classical ferromagnetic, antiferromagnetic,  frustrated states as well as the novel exotic states that can be associated with a spin wave, spin liquid, phase chaos and cross-breeds between them.  } 

\kk{The theoretical approach, presented here, is based on the well-known complex Ginzburg-Landau equation (cGLE) with a saturable nonlinearity  \citep{Wouters, Berloff}, that was successfully used for the modelling of polariton condensates. Written for the condensate wavefunction $\psi$ in one-dimension, cGLE reads as}
\begin{eqnarray}	
	i \hbar  \frac{\partial \psi}{\partial t} &=& - \frac{\hbar^2}{2m}\left(1 - i \eta_d N_R \right)\psi_{xx} + U_0 |\psi|^2 \psi+ 
	{\hbar g_R N_R(x)}\psi  \nonumber \\
	  &+&\frac{i\hbar}{2} \left( R_R N_R(x) - \gamma_C \right) \psi,
	\label{Initial_GL_equation}  
\end{eqnarray}
where $N_R = P(x)/(\gamma_R + R_R|\psi|^2)$ is the density of the hot exciton reservoir, $m$ is polariton effective mass, $U_0$ and  $g_R$ are the strengths of  effective polariton-polariton  and polariton-exciton interactions, respectively, $\eta_d$ is the energy relaxation coefficient specifying the rate at which gain decreases with increasing energy, $R_R$ is the rate at which the reservoir excitons enter the condensate, $\gamma_C$ and $\gamma_R$ are the rates of the condensate polaritons  and reservoir excitons  losses, respectively,  and $P$ is the  pumping  into the reservoir. When  pumped into several  spots with the outflows from each spot reaching its neighbours the system  establishes a global coherence with a chemical potential $\mu$ if the characteristics of the pump (intensity, spatial shape) are not vastly different from one spot to another. The steady state of such system satisfies
\begin{eqnarray}
\mu \Psi &=& -(1 - i\eta n)\Psi_{xx}+ |\Psi|^2 \Psi+ g n(x) \Psi +i \left( n(x)  -\gamma \right) \Psi, 
\label{Main} \\
n &=& \frac{p(x)}{(1+b|\Psi|^2)},
\label{Main_reservoir}
\end{eqnarray}
where we  non-dimensionalized Eq.~(\ref{Initial_GL_equation}) using 
$
        \psi  \rightarrow  \sqrt{\hbar^2  / 2m U_0 l^2} \Psi,
        {\bf r}  \rightarrow  l {\bf r}, t \rightarrow 2m t l^2/ \hbar$ and introducing the notations
  $g = 2 g_R/R_R,$ $\gamma = m \gamma_C l^2/ \hbar $,  $ p=m l^2 R_R P({\bf r})/ \hbar \gamma_R,\eta = \eta_d \hbar  / mR_R l^2,$ and
$ b = R_R \hbar^2  / 2m  l^2\gamma_R U_0. $ We choose the unit length as $l=1\mu m$. 
 The Madelung transformation $\Psi=\sqrt{\rho}\exp[i S]$ relates the wavefuntion to   density $\rho=|\Psi|^2$ and  velocity $u= S_x$.  To derive the coupling strength we consider a single pumping spot \nb{ centered at the origin and exponentially decaying to zero away from it}.   At large $x$, where $p(x)=0$, the velocity $u$ is given by the outflow wavenumber $k_c= const$ with $\rho_x/\rho=-\gamma/k_c,$ which can be integrated to yield $\rho\sim \exp[-x\gamma/k_c].$ From Eq.~(\ref{Main}), therefore, we obtain 
$
	\mu=k_c^2-\gamma^2/4k_c^2
$
at infinity. 
 %In Supplementary Materials we give the exact analytical solution of Eq. (\ref{Main}-\ref{Main_reservoir}) for some physically relevant pumping profiles and  approximate solutions for $p(x)=p_0 \exp(-\sigma x^2)$.  In both cases, the density profile has a form $\rho(x)=a_0/[\exp(\gamma x/k_c) + \xi - 2 + \exp(-\gamma x/k_c)]$, where $a_0=(4 k_c^4 \xi-\gamma ^2 \xi -4 \gamma ^2)/4k_c^2$ and $\xi$ is fixed by pumping and other system parameters. 
%
%
% We consider a 1D chain of $\ell$ identical condensates at positions $x_i$.  
%

\kk{
The wave function, that approximately describes the system of $\ell$ identical pumping spots at positions $x_i$, takes the form
$
 \Psi_g(x) \approx \sum_{i=1}^\ell \Psi_i (x-x_i),
 $
where the wave function of a single pumping spot centred at $x=x_i$ can be approximated by 
$
\Psi_i(x-x_i)=\sqrt{\rho(x-x_i)}\exp[({\rm i} k_c|x-x_i|)+{\rm i}\theta_i],
$
 where $\theta_i$ is a space independent part of the phase. In the expression for the space varying phase we neglected the healing of the outflow velocity to zero at the center of the pump. This healing occurs on the lengthscale of the order of the width of the pump and can be neglected when evaluating the integral quantities over the entire sample.  }
\kk{We note that nonlinearity of the system affects the assumption of the linear superposition of the  individual wavefunctions in several ways: not only it rescales the single wavefunction in the superposition and decreases the chemical potential of the superposition in the steady state, but also for small distances between the pumping spots may  bring about periodic and disordered fluctuations of the phase differences between the condensates. However, for well separated condensates this approximation is valid.}
 
 Depending on the pumping parameters, the system will lock with the relative phases $\theta_{ij}\equiv \theta_i-\theta_j$ between the sites $i$ and $j$ to achieve the highest
occupation number -- the total amount of matter given by 
${\cal D}=\int_{-\infty}^{\infty} |\Psi_g|^2 dx$  \cite{ohadi16}. 
To evaluate ${\cal D}$ we work in the Fourier space and  write %
 \begin{eqnarray}
 {\cal D}&=& \frac{1}{2 \pi}\int \large| \widehat{\Psi_g} (k)\large|^2 dk \approx \frac{1}{2 \pi}\int \large| \sum_i \widehat{\Psi}_i (k) \large|^2 dk = \nonumber \\
 &=& \ell {\cal D}_0 + \frac{1}{ \pi} \sum_{i<j } \int \left( \widehat{\Psi}_i \widehat{\Psi}_j^* + c.c. \right) dk
 \label{psi_fourier}
 \end{eqnarray}
where ${\cal D}_0$ is the number of particles of a single isolated condensate,  \nb{$\widehat{\Psi_g} (k)$ is the Fourier transform of $\Psi_g$,} and
\begin{eqnarray}
	\widehat{\Psi}_i(k) &=& \int_{-\infty}^{\infty} \Psi_i (x-x_i)  \exp(-{\rm i} k x)dx =  \nonumber \\
	&=& \exp( - {\rm i} k x_i) \int_{-\infty}^{\infty}  \Psi_i (\alpha) \exp(- {\rm i} k \alpha) d\alpha = \nonumber \\
	&=&  \exp( - {\rm i} k x_i + {\rm i} \theta_i) \widehat{\psi}(k) 
	\label{psi_i_k}, \\
	\widehat{\psi}(k) &=& 2\int_{0}^{\infty}  \sqrt{\rho (\alpha)} \exp({\rm i} k_c \alpha) \cos( k \alpha ) d\alpha.
\end{eqnarray}
Denoting the distances between the spots as   $x_{ij} = x_i - x_j$ we substitute (\ref{psi_i_k}) into the integral in Eq. (\ref{psi_fourier}) to get:
\begin{equation}
	{\cal D} = \ell{\cal D}_0
	+ \frac{2}{ \pi} \sum_{i<j}  \cos\theta_{ij} \int_0^{\infty}|\widehat{\psi}(k)|^2 \cos(k x_{ij}) dk.
	\label{ddd}
\end{equation}
This implies that during the condensation the system of pumping spots establishes the phase difference in such a way as to minimize ${\cal H}_{XY}$,
where the coupling strengths are given by
\begin{equation}
	J_{ij}= \frac{2}{ \pi}  \int_0^\infty |\widehat{\psi}(k)|^2 \cos(k x_{ij}) dk.
	\label{J_ij}
\end{equation}
%
%If $J_{ij}>0$ between two neighbours they establish a ferromagnetic coupling and in the case of two spots lock with a zero phase difference. If $J_{ij}<0$ two condensates coupled   antiferromagnetically, which leads to two spots to lock with   $\pi$ phase difference.

To obtain an analytical approximation of the coupling strengths and a criterion for the switching between the ferro- and antiferromagnetic connections along the chain of pumped condensate,  we parameterise the amplitude of the condensate by the inverse width $\beta$ and the height $A$: $\sqrt{\rho(x)}\approx A\exp[-\beta |x|]$. We expect the width and the height of the condensate   to  correlate with the width and the intensity of the pumping profile, respectively. For this approximation of the amplitude  the integrals in Eqs. (\ref{ddd}) and (\ref{J_ij}) can be evaluated exactly.
\begin{eqnarray}
\widehat{\psi}(k)&=&2 A \int_0^{\infty}  \exp(-\beta \alpha +{\rm i} k_c \alpha) \cos( k \alpha ) \, d\alpha\\
&=&\frac{2 A(\beta - i k_c)}{\beta^2 + k^2  - 2 i \beta k_c - k_c^2}.\nonumber\\
J_{ij}&=& \frac{8A^2}{ \pi}  \int_0^\infty \frac{(\beta^2 + k_c^2) \cos(k x_{ij})\, dk }{\beta^4 + (k^2-k_c^2)^2 + 2 \beta^2 (k^2+k_c^2)}. \label{jjj}
\end{eqnarray}
Applying the residue theorem for evaluating the last integral, we obtain the closed form expression for the coupling constants
\begin{equation}
J_{ij}=2A^2[\frac{1}{\beta} \cos(k_c x_{ij}) + \frac{1}{k_c} \sin(k_c x_{ij})] e^{-\beta x_{ij}}
.
\label{J_ijmain}
\end{equation}
This expression determines the switching of ferro- and antiferromagnetic coupling between the neigbours since the sign of $J_{ij}$ is set by the expression $ \cos(k_c x_{ij})/\beta + \sin(k_c x_{ij})/k_c$. If the pumping profile is wide ($\beta$ is small), the sign of the interactions is determined by $\cos(k_c x_{ij}),$ which is what we expect directly from Eq. (\ref{J_ij}) since $|\widehat{\psi}(k)|^2 \sim \delta(k - k_c)$ for a wide pumping spot. 

\begin{figure}[ht]
        \includegraphics[scale=0.3]{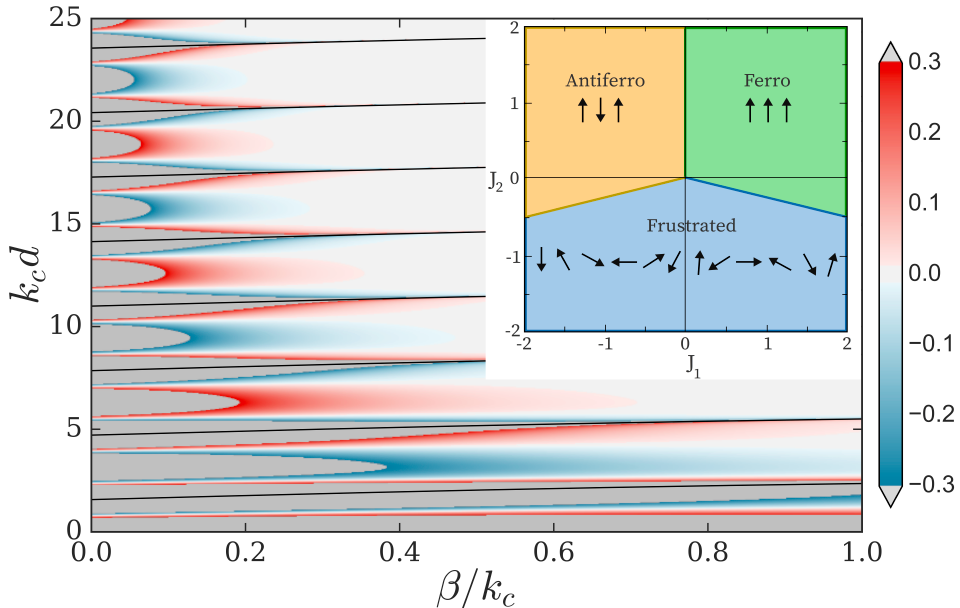}
	\centering
	\caption{\kk{Contour plot of frustration parameter $J_2/J_1$ as a function of $k_c d$ and $\beta/k_c$, where $k_c$ is the polariton outflow wavevector, $\beta$ is the inverse width of the polariton spot, $d$ is the distance between the two adjacent spots. The coupling strengths $J_1$ and $J_2$ are calculated from  Eq. (\ref{J_ijmain}) for the distances $d$ and $2d$, respectively.} The sign changes for   $J_1$ are shown by black solid lines. 
	%Non-resonant pumping profiles for realising different lattice configurations: linear chain by trapped condensates; (b) linear chain by condensates pumped into a circle. Black circles show the positions of the condensates. 
	The inset shows the different regimes of classical XY model on the $J_2-J_1$ plane: ferromagnetic, antiferromagnetic and frustrated.}
	\label{jj}
\end{figure}

Now we consider a  linear periodic chain of $\ell$ equidistant  polariton condensates separated by  $x_{ij}=d$.   
%On Fig. \ref{pump}(a,b) we show the schematics of the non-resonant pumping profiles for generating such a quasi-one-dimensional periodic chain. 
This chain can be achieved by creating a sequence of  trapped condensates  \cite{cristofolini, sun17}  or  by  pumping condensates around a circle \cite{natmat17,natphys17}. The corresponding XY model takes form ${\cal H}=-J_{1}\sum_{ i} {\bf s}_i\cdot{\bf s}_{i+1} -J_2 \sum_{ i} {\bf s}_i\cdot{\bf s}_{i+2}$, where the sum is over all $\ell$ condensates with periodic boundary conditions. In case of $J_2=0$ the model is integrable \cite{bethe}, whereas for $J_2\ne0$ the exact solutions were found for a  limited set of values of $J_2/J_1$. Frustrated  phases of the classical as well as quantum spin-$1/2$  system  with nearest-neighbour and next-nearest-neighbour isotropic exchange known as the Majumdar-Ghosh Hamiltonian have been extensively studied \cite{mg, bursill}. Classically, three regimes were identified for $\ell \rightarrow \infty$: ferromagnetic for $J_1>0,J_2>-J_1/4$, antiferromagnetic for  $J_1<0, J_2>J_1/4$ and frustrated (spiral) phase otherwise, as the inset to Fig. {\ref{jj} illustrates. In frustrated phase the pitch angle of the spiral is $\phi=\cos^{-1}(-J_1/4J_2)$ \cite{bursill}. Quantum fluctuations lower the ground state and shift the phase transition from spin liquid state at $J_2=0$ to a dimerized regime with a gap to the excited states at $J_2=0.2411 J_1$; the transition from antiferromagnetic phase to dimerized singlets takes place at $J_2=J_1/2$ \cite{mg}. In what follows, we elucidate if a polariton linear periodic chain is capable of reproducing the characteristics of classical regimes and what new physics arises due to nonlinear interactions of polaritons. 

 We plot the values  of $J_2/J_1$ found from Eq. (\ref{J_ijmain}) in Fig.~\ref{jj}   \kk{and demonstrate that for the experimentally achievable polariton spot widths and wavevectors the values of $J_2/J_1$ from $-0.3$ to $0.3$ should be realizable. } This suggests that all three of the classical regimes, \kk{that are depicted in the inset of the Fig.~\ref{jj}}, should be accessible \kk{in a linear periodic chain}.
 
 \bigskip

%\begin{figure}[ht]
%	\centering
%	\includegraphics[width=8.6cm]{Fig2,J2J1,v2.png}
%	\caption{Contour plot of $J_2/J_1$ as a function of $k_c d$ and $\sigma/k_c$. Lines of the sign change for   $J_1$ are shown by black solid lines. }
%	\label{jj}
%\end{figure}
%
\begin{figure}[ht]
	\includegraphics[width=8.6cm]{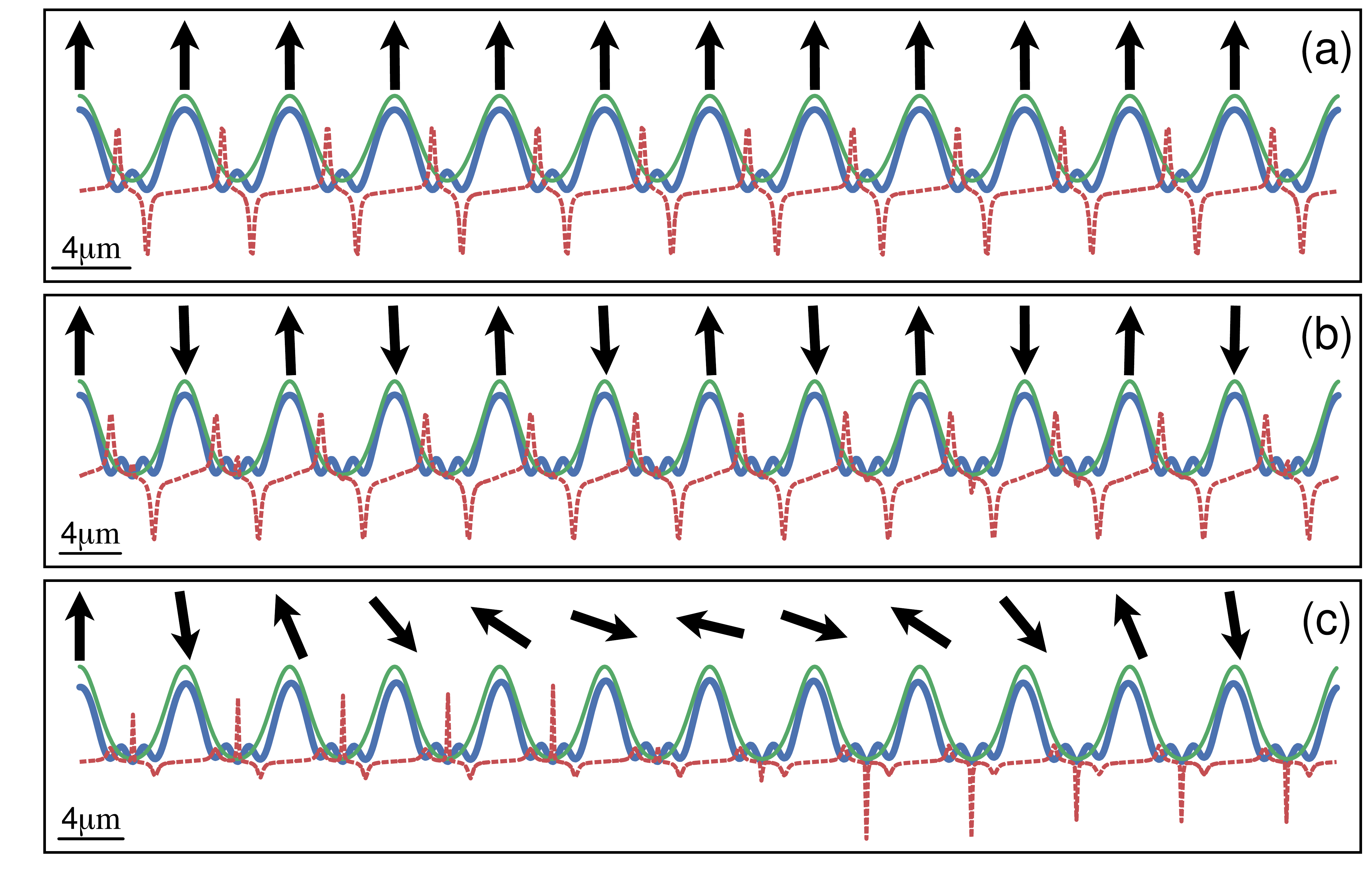}
	\centering
	\caption{Polariton densities (blue solid line) and velocities (red dashed lines)  for twelve condensates obtained by numerical integration of Eq.~(\ref{Initial_GL_equation}) with periodical boundary conditions. Pumping profiles are shown with green solid lines. Panel  (a) shows the ferromagnetic state,  (b) corresponds to antiferromagnetic state with $\pi$ phase difference between the adjacent sites, and the panel (c) shows the frustrated state. \kk{The distances between the nearest condensates are $d = 5.4 \mu m$, $d = 6.9 \mu m$, and $d = 6.5 \mu m$ for (a), (b), and (c), respectively. The numerical parameters for 1D simulations are $\eta = 0.4$, $b = 1.5$, $\gamma = 1$, $p_0 = 5$, $\sigma = 0.4$, $g = 2.5$.}}
	\label{1d}
\end{figure}
%
%\begin{figure}[ht]
%	\includegraphics[width=8.6cm]{Fig3,1D_period_v2.png}
%	\centering
%	\caption{The twelve condensates collapse into six condensates and show oscillating frustrated state.}
%	\label{1d,period}
%\end{figure}
%
\begin{figure}[ht]
	\includegraphics[scale=0.4]{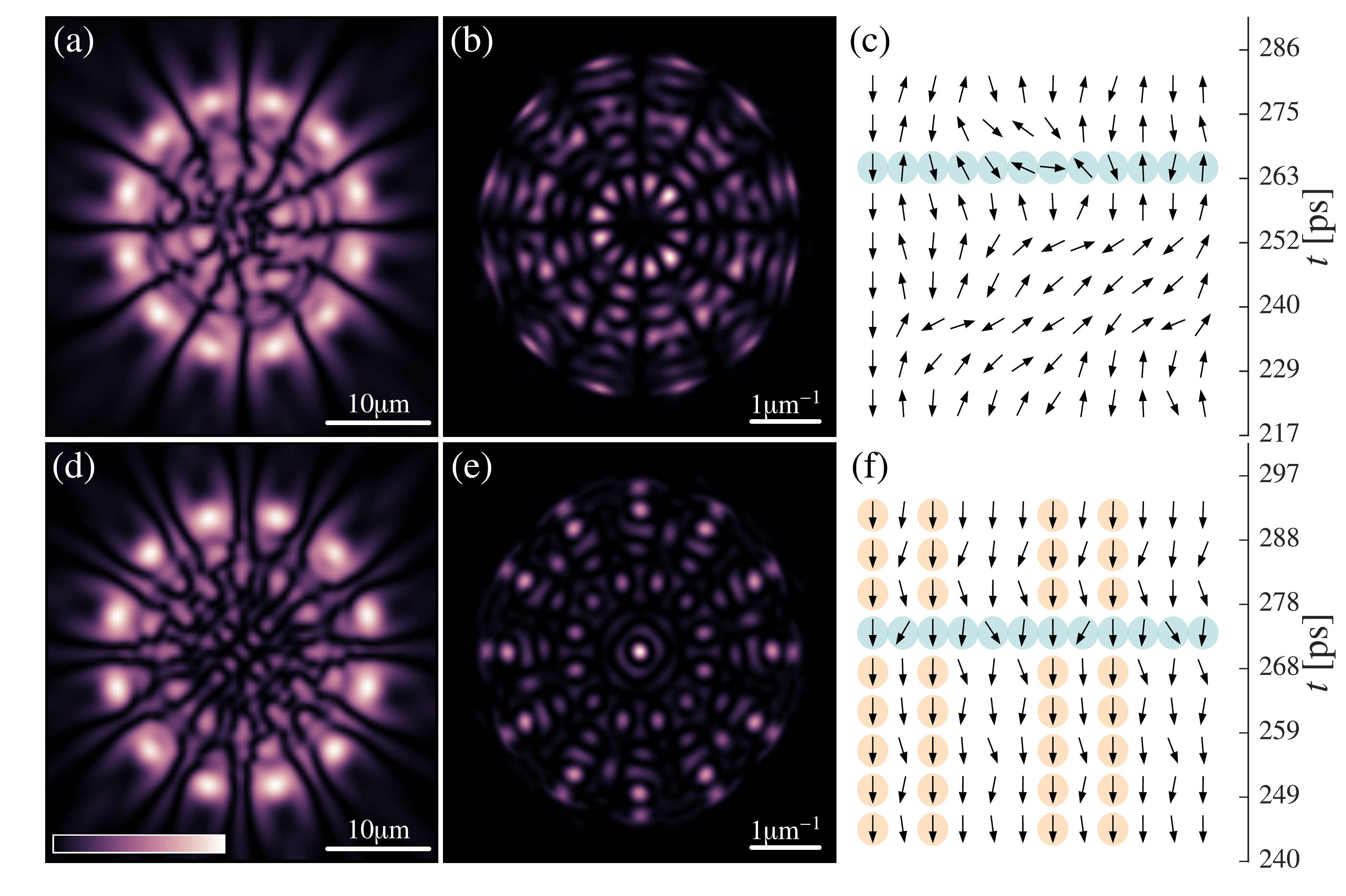}
	\centering
	\caption{ \kk{Density snapshots (left panel) and the far field emission  (central panel) at a fixed time for twelve condensates arranged in a circle obtained by numerical integration of Eq.~(\ref{Initial_GL_equation}). The right panels show the evolution of the phases (relative to one fixed spin) between adjacent spots with 12$ps$ (c) and 10$ps$ (f) time laps between the spin configurations. The highlighted with blue circles spin configurations in (c) and (f) correspond to the density profiles (a,d) and the far field emissions (b,e). The steady spins of the spots are marked with orange circles in (c,f).} We use the simulation parameters that were found to agree with experimental data in our previous works \cite{natmat17, param}, except for the pumping intensity which is 2.6 times larger to bring about a non-stationary state; \kk{the distances between the adjacent sites are $6.4\mu m$ (a-c) and $7\mu m$ (d-f)}.}
	\label{2d}
\end{figure}

Next  we consider twelve  condensates in a linear periodic chain and numerically integrate  Eq.~(\ref{Initial_GL_equation})  for a variety of distances between the spots. For each configuration we start from 100 random initial distribution of phases to find the ground state configurations. Figure \ref{1d} identifies ferromagnetic, antiferromagnetic and spiral spin wave phases that represent the ground states of the one-dimensional XY model. \kk{The polariton densities (blue solid lines) are clearly displaced from the pumping profiles (green solid lines) in case of the spin wave state which is depicted in Fig.~\ref{1d}(c). This state has a distinguished velocity pattern (red dashed lines) compared to the other two classical states in Fig.~\ref{1d}(a,b). The experimental systems may suffer from noise, sample disorder, interactions with impurities, so we have repeated the numerical simulations with the Langevin noise described in our previous work \cite{ohadi16}. This noise introduces small oscillations around the steady states, but has no effect on the time averaged solutions that coincide with those found without the Langevin noise term. } 

We have also considered the full two-dimensional system \nb{with pumping spots equidistant around a circle.} For pumping intensities just above the threshold the same classical phases are obtained. As the pumping intensity increases, the nonlinearity of the system destabilises the frustrated state and produces spin fluctuations, as Fig.\ref{2d}(a-c) illustrates, suggesting the formation of \kk{a non-stationary and chaotic spin wave which could be related to phase chaos \cite{phasechaos}.}  
\kk{Direct observation of non-stationary states in experiments would require time-resolved measurements on time scales that are challenging with current experimental configurations. Note, that this is the lowest energy state with respect to the XY Hamiltonian, as it carries the largest number of particles for the given network configuration. It suggests that the pumping intensity can be used to continuously move between various state configurations: frustrated states, spin waves, spin liquids, phase chaos etc. and to study transitions between them.} 
\kk{For instance, Fig. \ref{2d}(d-f) demonstrates a non-stationary state of two spin waves of different periods (one or three spins). The spectral weights at a fixed time (Fig. \ref{2d}(b,e)) reflect the symmetry of the lattice.}

\kk{Non-stationary spin wave states} can be detected in the momentum- and energy-resolved photoluminescence spectrum, which can be directly measured in the far field. Figure \ref{Fourier} shows the spectral weight, $I(\omega, {\bf k})=\biggl|\iint \Psi({\bf r}, t) \exp[-i{\bf k}\cdot {\bf r}-i\omega t]dtd{\bf r}\biggr|^2$,  as a function of  $(\omega, k_x, k_y=0)$. \kk{In the case of the non-stationary state depicted in Fig. \ref{2d}(a-c), spins reorient themselves randomly with time, cycling through different micro-states  with the density distribution depicted in Fig. \ref{Fourier}(a). The state shown in Fig. \ref{2d}(d-f) is a more regular state  with  the energy spectrum  shown in Fig. \ref{Fourier}(b) indicating  several well separated energy levels. The stationary state  would, in contrast, show only one energy level \cite{borgh1,borgh2}. }

Frustrated states that we found in the linear periodic chain of polariton graphs correspond to  superfluids at nonzero quasi-momentum and, therefore, exhibit  nontrivial long range
phase order. The spiral phases  spontaneously break time-reversal symmetry by generating   bosonic currents around the sites of polariton graphs. %\cite{struck11,shmied08}.

\begin{figure}[ht]
	\includegraphics[scale=0.4]{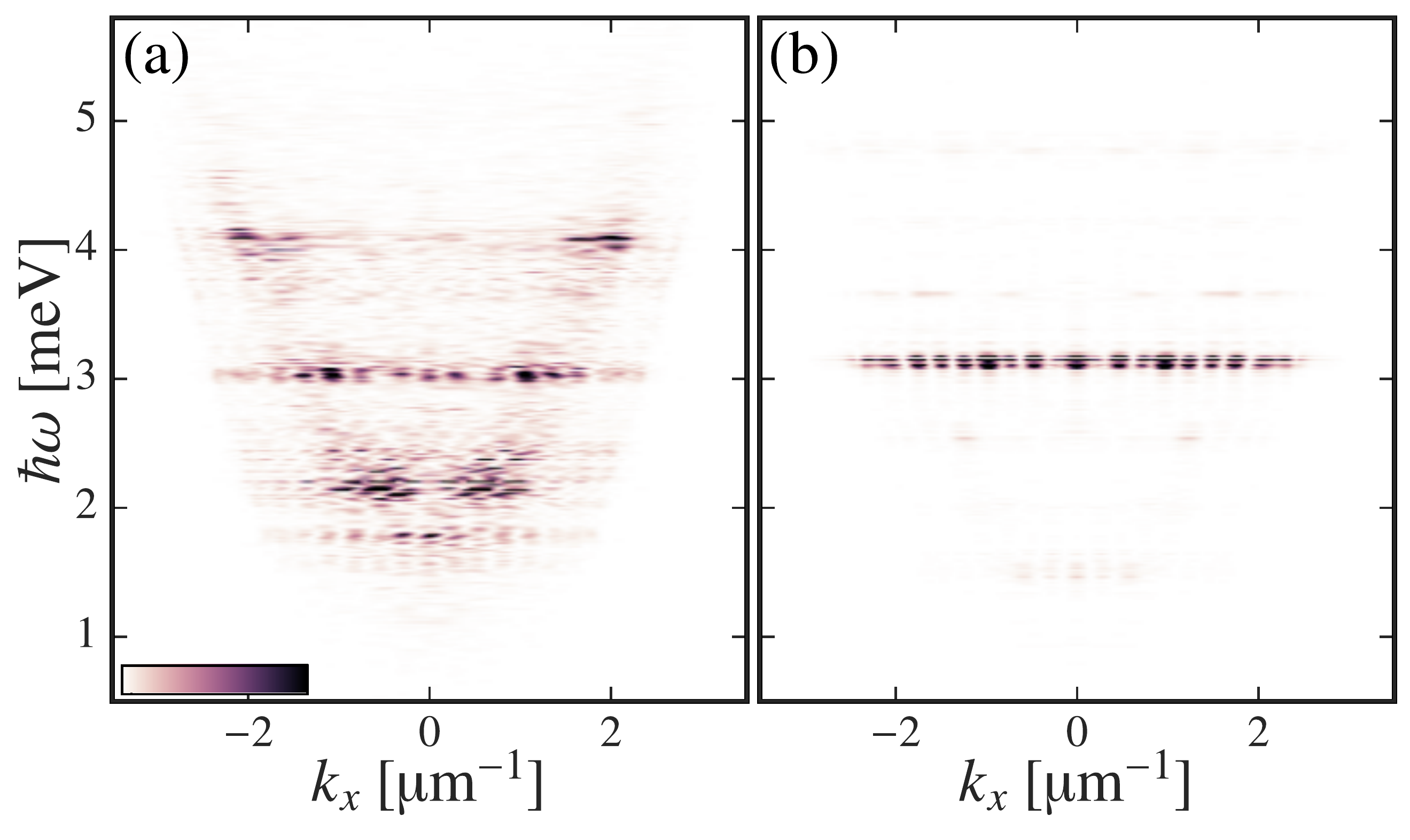}
	\centering
	\caption{ The spectral weight plots for $k_y=0$ of the two-dimensional non-stationary spin wave states from Fig. \ref{2d}(a-c) in (a) and from Fig. \ref{2d}(d-f) in (b). \kk{Both plots are saturated at the same level of $0.2$ to make the energy levels more visible and to provide an easier comparison between the states}. }
	\label{Fourier}
\end{figure}
In conclusion, we considered   polariton condensates arranged in a periodic linear chain. We evaluated the interaction strength between the condensates analytically in terms of the outflow wavevector, pumping width and strength, and the distance between the condensates. We have identified parameter regime where the interactions beyond the next neighbours become important and lead to the appearance of the classical frustrated state. The nonlinear interactions beyond the linear approximation lift the ground state degeneracy and facilitate the formation of a disordered state.  We demonstrated that a sequence of  polariton condensates  are capable of supporting not only classical ground states of the XY model with nearest and next to nearest neighbour interactions, but also exotic states, such as \kk{non-stationary spin waves that may be associated with spin liquids or phase chaos}. This observation opens a new path to studying such novel states of matter.

%%%%%%%%%%%%%%%%%%%%%%%%%%

%%%%%%%%%%%%%%%%%%%%%%%%%%%


\begin{thebibliography}{00}
%
\bibitem{Sachdev_NatPhys2008} S. Sachdev, Quantum magnetism and criticality, \textit{Nat. Phys.} \textbf{4}, 173 (2008).
%
\bibitem{spins} L. Balents, Spin liquids in frustrated magnets, \textit{Nature} \textbf{464}, 199 (2010). 
%
\bibitem{hubbard} F. H. L. Essler, H. Frahm, F. Gohmann, A. Klumper, and V. E. Korepin, The one-dimensional Hubbard Model (Cambridge University Press, Cambridge, 2005).
%
\bibitem{Auerbach1994} A. Auerbach, Interacting electrons and quantum magnetism (Springer, New York, 1994).
%
\bibitem{spinliquids} G. Misguich and C. Lhuillier,  Frustrated spin systems, edited by H.T. Diep (World Scientific, Singapore, 2004). 
%
\bibitem{Alet2006} F. Alet, A. M. Walczak, and M. P. A. Fisher, Exotic quantum phases and phase transitions in correlated matter, \textit{Physica A} \textbf{369(1)}, 122-142 (2006).
%
\bibitem{balents} L. Balents, Spin liquids in frustrated magnets, \textit{Nature} \textbf{464}, 199-208 (2010).
%
\bibitem{anderson} P. W. Anderson, Resonating valence bonds: A new kind of insulator? \textit{Materials Research Bulletin} {\bf 8(2)}, 153-160 (1973).
%
\bibitem{han} T.-H. Han, J. S. Helton, Sh. Chu, D. G. Nocera, J. A. Rodriguez-Rivera, C. Broholm, and Y. S. Lee, Fractionalized excitations in the spin-liquid state of a kagome-lattice antiferromagnet, \textit{Nature} \textbf{492}, 7429 (2012).

\bibitem{reviewUltracold}  M. Lewenstein, A. Sanpera, V. Ahufinger, B. Damski, A. Sen, and U. Sen, Ultracold atomic gases in optical lattices: mimicking condensed matter physics and beyond, \textit{Advances in Physics} \textbf{56(2)}, 243-379 (2007).

\bibitem{saffman}  M. Saffman, T. G. Walker, and K. Molmer, Quantum information with Rydberg atoms. \textit{Rev. Mod. Phys.} \textbf{82}, 2313 (2010).

\bibitem{struck11} J. Struck, C. Ölschläger, R. Le Targat, P. Soltan-Panahi, A. Eckardt, M. Lewenstein, P. Windpassinger, and K. Sengstock, Quantum simulation of frustrated classical magnetism in triangular optical lattices, \textit{Science} {\bf 333}, 996 (2011).
%
\bibitem{simon11} J. Simon, W. S. Bakr, R. Ma, M. E. Tai, Ph. M. Preiss, and M. Greiner, Quantum simulation of antiferromagnetic spin chains in an optical lattice, \textit{Nature} \textbf{472}, 307-312 (2011).
%
\bibitem{pla} J. J. Pla, K.Y. Tan, J. P. Dehollain, W. H. Lim, J. JL Morton, F. A. Zwanenburg, D. N. Jamieson, A. S. Dzurak, and A. Morello, High-fidelity readout and control of a nuclear spin qubit in silicon, \textit{Nature} \textbf{496}, 334-338 (2013).
%
\bibitem{hanson} R. Hanson and D. D. Awschalom, Coherent manipulation of single spins in semiconductors, \textit{Nature} \textbf{453}, 1043-1049 (2008).
%
\bibitem{lemeshko13} M. Lemeshko, N. Y. Yao, A. V. Gorshkov, H.Weimer, S. D. Bennett, T. Momose, and S. Gopalakrishnan, \textit{Phys. Rev. B} {\bf 88}, 014426 (2013).
%
\bibitem{northup} T. E. Northup and R. Blatt, Quantum information transfer using photons, \textit{Nature Photonics} \textbf{8}, 356 (2014).
%
\bibitem{lanyon} B. P. Lanyon, C. Hempel, D. Nigg, M. Müller, R. Gerritsma, F. Zähringer, P. Schindler et al. Universal digital quantum simulation with trapped ions, \textit{Science} \textbf{334}, 57 (2011).
%
\bibitem{kim10} K. Kim, M-S. Chang, S. Korenblit, R. Islam, E. E. Edwards, J. K. Freericks, G-D. Lin, L-M. Duan, and C. Monroe. Quantum simulation of frustrated Ising spins with trapped ions, \textit{Nature} \textbf{465}, 590-593 (2010).
%
\bibitem{corcoles} A. D. Corcoles  et al. Demonstration of a quantum error detection code using a square lattice of four superconducting qubits, \textit{Nature Commun.} \textbf{6}, 6979 (2015).
%
\bibitem{yamamoto11} S. Utsunomiya, K. Takata and Y. Yamamoto,  Mapping of Ising models onto injection-locked laser systems, \textit{Opt. Express} \textbf{19}, 18091 (2011).
%
\bibitem{yamamoto14} A. Marandi, Z. Wang, K. Takata, R. L. Byer, and Y. Yamamoto,  Network of time-multiplexed optical parametric oscillators as a coherent Ising machine, \textit{Nature Photonics} \textbf{8}, 937-942 (2014).
%
\bibitem{coupledlaser} M. Nixon, E. Ronen, A. A. Friesem, and N. Davidson, Observing geometric frustration with thousands of coupled lasers, \textit{Phys. Rev. Lett.} \textbf{110}, 184102 (2013).
%
\bibitem{natmat17}  N. G. Berloff,	M. Silva,	K. Kalinin,	 A. Askitopoulos,	J. D. T\"opfer,	P. Cilibrizzi,	W. Langbein	and P. G. Lagoudakis, Realizing the classical XY Hamiltonian in polariton simulators, {\it Nature Materials} doi:10.1038/nmat4971 (2017)
\bibitem{tosi12} G. Tosi, G. Christmann, N. G. Berloff, P. Tsotsis, T. Gao, Z. Hatzopoulos, P. G. Savvidis, and J. J. Baumberg, Sculpting oscillators with light within a nonlinear quantum fluid, \textit{Nature Physics} \textbf{8}, 190-194 (2012).
%
\bibitem{ohadi16} H. Ohadi, R. L. Gregory, T. Freegarde, Y. G. Rubo, A. V. Kavokin, N. G. Berloff, and P. G. Lagoudakis, Nontrivial Phase Coupling in Polariton Multiplets, \textit{Phys. Rev. X} {\bf  6}, 031032 (2016).
%
\bibitem{Wouters}
M. Wouters and I. Carusotto, Excitations in a nonequilibrium Bose-Einstein condensate of exciton polaritons, \textit{Phys. Rev. Lett.} \textbf{99}, 140402 (2007).
%
\bibitem{Berloff}J. Keeling and N. G. Berloff, Spontaneous rotating vortex lattices in a pumped decaying condensate, \textit{Phys. Rev. Lett.} {\bf 100}, 250401 (2008).
%
\bibitem{cristofolini} P. Cristofolini, A. Dreismann, G. Christmann, G. Franchetti, N.G. Berloff, P. Tsotsis, Z. Hatzopoulos, P.G. Savvidis, J.J. Baumberg, Optical superfluid phase transitions and trapping of polariton condensates, \textit{Physical Review Letters} \textbf{110}, 186403 (2013).
%
\bibitem{sun17} Y. Sun, P. Wen, Y. Yoon, G.Liu, M. Steger, L. N. Pfeiffer, K. West, D. W. Snoke, and K. A. Nelson, \textit{Phys. Rev. Lett.} {\bf 118}, 016602 (2017).

\bibitem{natphys17} K. Kalinin, M.  Silva, W. Langbein, N.G. Berloff, P.G. Lagoudakis, Spontaneous discrete vortex solitons in polariton lattices, submitted to Nature Physics (2017).

%\bibitem{param} The dimensionless parameters used in 1D simulations are: $\eta = 0.4$, $b = 1.5$, $\gamma = 1$, $P_0 = 5$, $\sigma = 0.4$, $g = 2.5$.
%
\bibitem{bethe} H. Bethe, "Zur Theorie der Metalle. I. Eigenwerte und Eigenfunktionen der linearen Atomkette". (On the theory of metals. I. Eigenvalues and eigenfunctions of the linear atom chain), Zeitschrift für Physik, 71:205226 (1931).

\bibitem{mg} C. K. Majumdar  and D. K. Ghosh, \textit{J. Math Phys.} {\bf 10}, 1388; {\bf 10} 1399, (1969). 

\bibitem{bursill} R. Bursill, G. A. Gehring, D. J. J. Farnell, J. B. Parkinson, Tao Xiang, and Chen Zeng, Numerical and approximate analytical results for the frustrated spin-1/2 quantum spin chain, \textit{J. Phys.: Condens. Matter 7} \textbf{45}, 8605-8618 (1995).

\bibitem{phasechaos} O. V. Popovych, Yu. L. Maistrenko, and P. A. Tass, Phase chaos in coupled oscillators, {\it Phys. Rev. E} {\bf 71}, 065201(R) (2005))
%
%\bibitem{qsl} Yi Zhou, K. Kanoda, and Tai-Kai Ng, Quantum Spin Liquid States, arXiv:1607.03228 (2016).
%
%\bibitem{schulz} H. J. Schulz, T. A. L. Ziman, and D. Poilblanc, Magnetic systems with competing interactions: frustrated spin systems, ed. H. T. Diep, World Scientific Publishing (1994).
%
%\bibitem{struck11} J. Struck, C. Ölschläger, R. Le Targat, P. Soltan-Panahi, A. Eckardt, M. Lewenstein, P. Windpassinger, and K. Sengstock, Quantum simulation of frustrated classical magnetism in triangular optical lattices, \textit{Science} {\bf 333}, 996 (2011).
%
%\bibitem{shmied08} R. Schmied, T. Roscilde, V. Murg, D. Porras, J. I. Cirac, \textit{N. J. Phys.} \textbf{10}, 045017 (2008).


\bibitem{borgh1} M. O. Borgh, J. Keeling and N.G. Berloff, Spatial pattern formation and polarization dynamics of a nonequilibrium spinor polariton condensate, \textit{Phys. Rev. B} {\bf 81}, 235302, (2010)



\bibitem{borgh2} M. O. Borgh, G. Franchetti, J. Keeling and N. G. Berloff, Robustness and observability of rotating vortex-lattices in an exciton-polariton condensate, \textit{Phys. Rev. B} {\bf  86}, 035307 (2012)

\end{thebibliography}
\end{document}